\documentclass[epjST]{svjour}
\usepackage{graphics,epsfig}
\begin{document}
\title{Analytic properties \\ of different unitarization schemes }
\author{O. V. Selyugin\inst{1}\fnmsep\thanks{\email{selugin@theor.jinr.ru}}
\and J.-R. Cudell\inst{2}\fnmsep\thanks{\email{JR.Cudell@ulg.ac.be}}
 \and E. Predazzi\inst{3}\fnmsep\thanks{\email{predazzi@to.infn.it}}
}
\institute{ BLTP, Joint Institute for Nuclear Research,
141980 Dubna, Moscow region, Russia \and AGO dept., Universit\'e de Li\`{e}ge, Belgium  \and Torino 
University,
Italy and INFN, Torino, Italy 
}
\abstract{
 The analytic properties of the eikonal and $U$-matrix unitarization 
sche\-mes are examined. It is shown that 
the basic properties of these schemes
are identical. Both can fill
the full circle of unitarity, and both can lead to
 standard and non-standard asymptotic relations for 
$\sigma_{el}/\sigma_{tot}$.
The relation between the phases of the unitarised amplitudes in each 
scheme is examined, and it is shown that
demanding equivalence of the two schemes leads to a bound on the
phase in the $U$-matrix scheme.
} 

\maketitle
\section{Introduction}
At the LHC,  the investigation of diffraction processes will occupy an 
important place. However, the diffraction  processes at  very high
energies  does not simplify asymptotically, but can display complicated 
features    \cite{FS-07,dif04}. This concerns especially the asymptotic 
unitarity bound connected with the so-called Black Disk Limit (BDL).

Summation of different sets of diagrams in the tree approximation can
lead to different unitarization procedures of the Born scattering 
amplitude. In the partial-wave language, we need to sum many different 
waves with $l \rightarrow \infty $ and this leads to the impact parameter 
representation~\cite{Predazzi66} converting the summation over $l$  into an 
integration over $b$.

The different unitarization procedures are then naturally formulated in the 
impact-parame\-ter representation. As they include 
different sets of inelastic states in the $s$ channel,
this leads to  unitarization schemes with  different  coefficients in front 
of the multiple exchanges. 

In the  impact parameter representation, we shall write the  Born term of 
the amplitude as
\begin{eqnarray}
 \chi(s,b) \ =  \frac{1}{4 \pi}
   \ \int_{0}^{\infty} \  e^{i \vec{b}\cdot \vec{q} } \  F_{Born}(s,q^2)
  \ d^2 \vec{q},
\label{tot0}
\end{eqnarray}
 where we have dropped the kinematical factor $1/\sqrt{s(s-2m_p^2)}$ and a 
factor $s$ in front of $F$.
For proton-proton scattering, where five helicity amplitudes exist,
the non-flip Born term in the near forward direction is 
$F_{Born}(s,t)=F^{+}_{Born}(s,t) = F_1(s,t)+F_3(s,t)$.
and $\chi(s,b)$ is, in first approximation, given by that non-flip helicity  
amplitude only.
  After the unitarization procedure we get
\begin{eqnarray}
 \sigma_{tot}(s)  &=& \ 4 \pi 
    \ \int_{0}^{\infty}   G( \chi(s,b))  \ b  \ db.
\label{overlap}
\end{eqnarray}

   Two unitarization schemes are more commonly used: one is the eikonal
   representation where

 \begin{eqnarray}
  G(s,b)  =  [1 \ - \ e^{-  \chi(s,b)}],
\label{eik1}
\end{eqnarray}
 and the  other  is the $U$-matrix  representation
\cite{bg,Chrustalev} where
\begin{eqnarray}
  G(s,b)  = U_T(s,b) = 2 \frac{ \hat{\chi}(s,b)}{1 \ + \ \hat{\chi}},
\label{U-mt}
\end{eqnarray}
with
$\hat\chi(s,b)=\chi(s,b)/2$. Of course, the eikonal representation can be 
obtained in different ways starting from simple diagrams in the tree 
approximation. But many additional diagrams exist when inelastic states in 
the s-channel are taken into account. Other approaches exist, see for 
example \cite{Martiros,Kaidalov1,Martynov1}, in which renormalized eikonal
representations were obtained. No one really knows which are the
leading diagrams and how to sum them. Therefore, all these
approaches of the eikonal scheme remain phenomenological.

There are two most important conditions which must be satisfied by any 
unitarization scheme. Firstly, in the limit of small energies, every 
unitarization representation must reduce to the same scattering amplitude.
Of course, the way this happens can depend on the representation of the 
Born amplitude in the model. The various unitarization schemes will give 
different results \cite{CS-sat06,CS-BDL06} only at high energies, when the 
number of diagrams and their forms are essentially different.
Secondly, the unitarized amplitude cannot exceed the upper unitarity bound. 
In the different normalizations, this bound may equal to $1$ or $2$.

At LHC energies we will be in a regime close to the unitarity bound. Hence,
it is very important to specify the possible domains of validity of the 
various  unitarization schemes and to determine, from the analysis of the  
experimental data, which form really corresponds to the physical picture.

\section{Properties of the standard $U$-matrix approach}
The standard $U$ matrix was intensively  explored in \cite{Chrustalev},
where it was written, in the partial-wave language, as
$U_{l}(t)$ so that it is bounded by  $U_{l} \leq 1$. In the impact 
parameter representation,  the properties of the $U$ matrix were explored in
\cite{TT1}, where  the $U$ matrix  was taken as pure imaginary. 
We will  denote it  by $i U_{T}(s,b)$ and it corresponds
to the following definition of the Born amplitude, which varies
from 0 to $\infty$:
\begin{eqnarray}
\hat{\chi}(s,b)\ = \ \frac{1}{2}\Im m (F^{Born}(s,b)).
\end{eqnarray}
In this case, the relation with the $S$ matrix will be
\begin{eqnarray}
S(s,b) = \frac{1 - \hat\chi(s,b)}{1 + \hat\chi(s,b)}.
 \label{S-U}
\end{eqnarray}
We can see that when $\hat\chi$ varies in the domain $ [0, \infty ] $, 
the $S$ matrix varies in the interval $[-1,+1]$. Hence the
scattering amplitude in the impact parameter space
$G(s,b)=T(s,b)= 1-S(s,b)$ will span the domain  $[0,2]$. The upper
value is the maximum of the unitarity bound. In this case, the
unitarised amplitude can fill the full circle of unitarity, and remains in 
it for all energies.

This form of unitarization leads to unusual properties at super-high
energies as was shown in \cite{TT1}; with non-standard properties as $ s 
\rightarrow  \infty$:
$\sigma_{inel}/\sigma_{tot} \rightarrow \ 0$,
$\sigma_{el}/\sigma_{tot} \rightarrow 1$.

In a recent paper \cite{TT07}, it has been claimed that such properties 
arise from the difference between the rational representation ($U$ 
matrix) 
and the exponential representation (eikonal). Let us show that it is not 
so.  We can consider an extended eikonal in the form
\begin{eqnarray}
\sigma_{tot}(s)  &=& \ 8 \pi  
\ \int_{0}^{\infty}   E_{R}(s,b)\ b \ db \ = \ 8 \pi
    \ \int_{0}^{\infty}   [1 \ - \ e^{- \hat{\chi}(s,b)}]  \ b\ db.
\label{eik2}
\end{eqnarray}
Formally, this is the same as the standard form of the eikonal
(\ref{eik1}),  except for the coefficient in front of the overlap
function and  for the use of $\hat{\chi}(s,b)$  instead of
$\chi(s,b)$. At small energy (i.e. small
 $\hat{\chi}$), we obtain the standard Born amplitude.
But  at high energy, we obtain the same  properties as in the case of the
 $U_T$-matrix representation. Indeed, the inelastic cross section can
be written
\begin{eqnarray}
    \eta(s,b)  =  
 \frac{1}{2} \  \exp(- \hat{\chi}(s,b)) \ [1  - \exp(- \hat{\chi}(s,b))] ,
\end{eqnarray}
which leads to   the same  result as the standard $U_T$ matrix \cite{TT1}:
$\sigma_{inel}/\sigma_{tot} \rightarrow 0$ and $\sigma_{el}/\sigma_{tot} \rightarrow 
1$ as $ s \rightarrow  \infty$.

 Our calculations
 for the inelasticity corresponding to $U_T$ and  $E_{R}$  are shown in 
Fig. 1. We see that  both solutions have the same behavior in $s$ and $b$
but $E_{R}$  has  sharper anti-shadowing properties. Both solutions lie in 
the unitarity circle, reach the maximum of the unitarity bound as $s 
\rightarrow \infty$, and have the same analytic properties asymptotically.

%
\begin{figure}[!ht]
\begin{center}
\epsfysize=50mm
\epsfbox{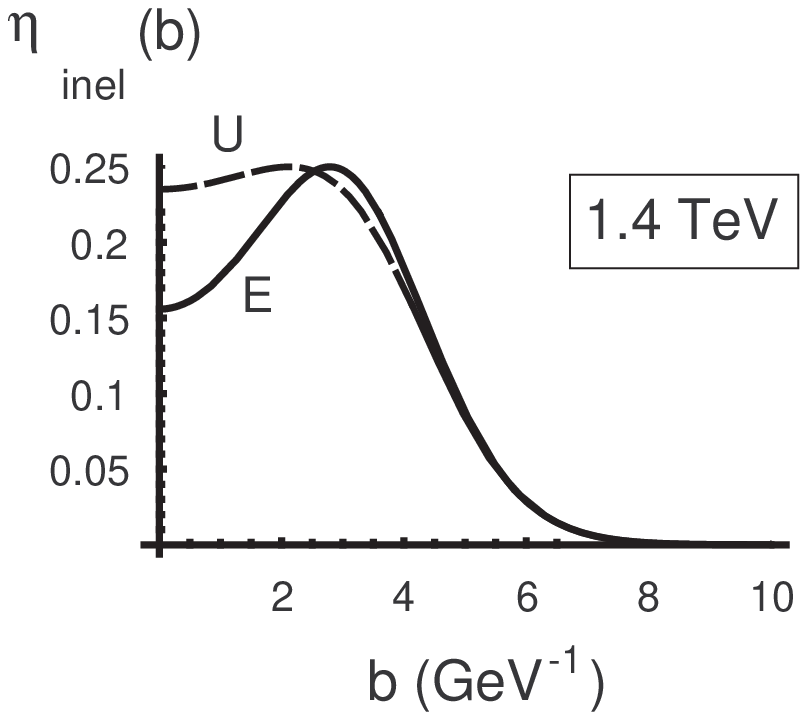}
\epsfysize=50mm
\epsfbox{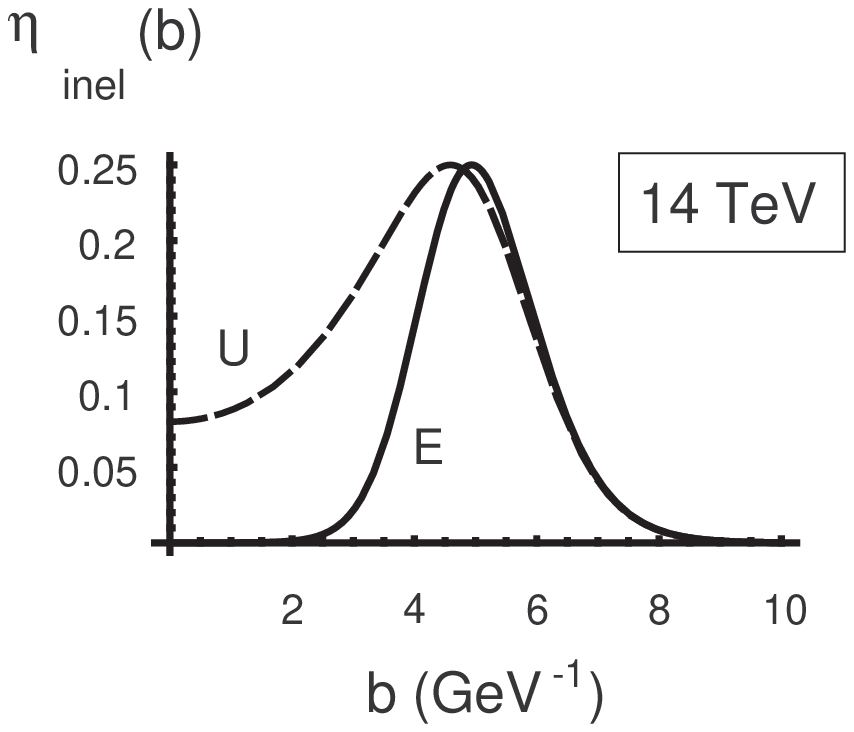}
\end{center}
\caption{  Antishadowing effects with eikonal (full line) and
 $U_{T}$ matrix (dashed line) $\sqrt{s}=1.4 \ $TeV (left) and
     $\sqrt{s}=14 \ $TeV  (right)
[$U_{T}(s,b)= \hat{\chi(s,b)} = 0.005 \ s^{0.4} \ exp(-b^2/3^2) $]
}
\end{figure}

  \begin{figure}
\vglue -1cm\epsfysize=70mm
\centerline{
\epsfbox{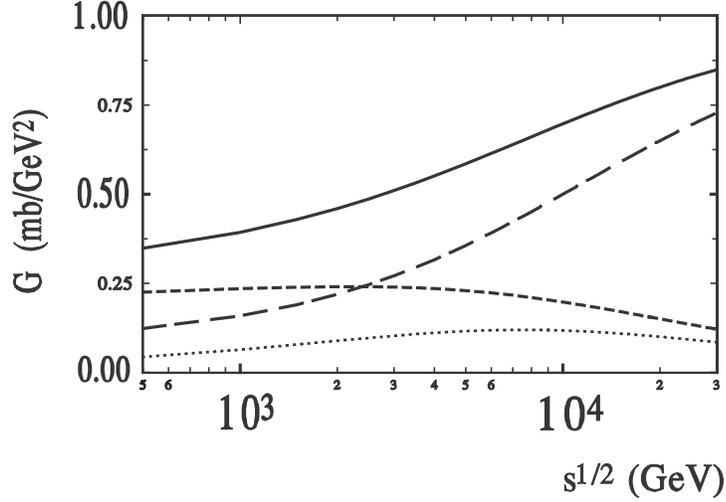}
}
\caption{ The energy dependence of the amplitudes from $U_{T}(s,b=0)$ 
for a model with a soft and a hard pomerons
 (the full and dotted lines are the imaginary part and the modulus of the 
real part, the long dashed line gives $G_{el}(s,b=0)$, and the dotted line 
is $G_{inel}(s,b=0)$). }
\end{figure}
Hence, non-standard analytic asymptotic properties 
are  not unique to the $U_T$ matrix with its special form
(\ref{U-mt}), and can be reproduced by
the exponential form (\ref{eik2}) for  $E_{R}$.

Most importantly, at LHC  energies,  these forms of unitarization do no 
have a Black-Disk limit. Indeed, the amplitude does not reach
 the saturation regime and the basic parameters of the scattering 
amplitude, such  as $\rho(s,t)$ and the slope $B(s,t)$,
do not change their behavior (see Fig. 2). It is only at still larger 
energies, when $\eta_{inel}(s,b)$ goes to zero
and the scattering amplitude reaches the unitarity limit,
that $\rho(s,t)$ and $B(s,t)$ exhibit properties similar to those
in the saturation regime.

\section{Properties of the standard eikonal and comparison with the $U_T$
matrix}
  In order to extend the $U$ matrix, and make it similar to the eikonal, we 
find it more transparent to work in a different normalisation, where 
$\sigma_{tot}(s) = 4 \pi Im F(s,t)$. The standard form of eikonal does not 
change, but we take an extended $U$-matrix unitarization $U_e$ with an 
additional coefficient $1/2$ in the denominator, as
\begin{eqnarray}
 \sigma_{tot}(s) & = & \ 4 \pi 
     \  \int_{0}^{\infty}  b \
          [\frac{\chi(s,b)}{(1 \ + \ \chi(s,b)}] \ db.   \label{Ue}
\label{um1b}
\end{eqnarray}
This form of the $U$ matrix then satisfies all the analytical properties of 
the standard eikonal representation: the inelastic overlap function  is
  \begin{eqnarray}
    \eta(s,b) \ = \ \frac{1}{2} \frac{ \chi(s,b) + 2 \chi^2(s,b)}{[1+ 
\chi(s,b))]^2}, \nonumber
\end{eqnarray}
and it  can easily be  seen that in this case,
   when  $ s \rightarrow  \infty$,  we obtain:
  $\sigma_{inel} \rightarrow \ \sigma_{el}$ and
   $\sigma_{el}/\sigma_{tot} \rightarrow 1/2$.
Differently stated, this $U_{e}$-matrix representation has
the standard BDL.
   In Figs. 3,  $\eta(s,b)_{inel}$  is shown  for these cases.
%
\begin{figure}[!ht]
\begin{center}
\epsfysize=50mm\epsfbox{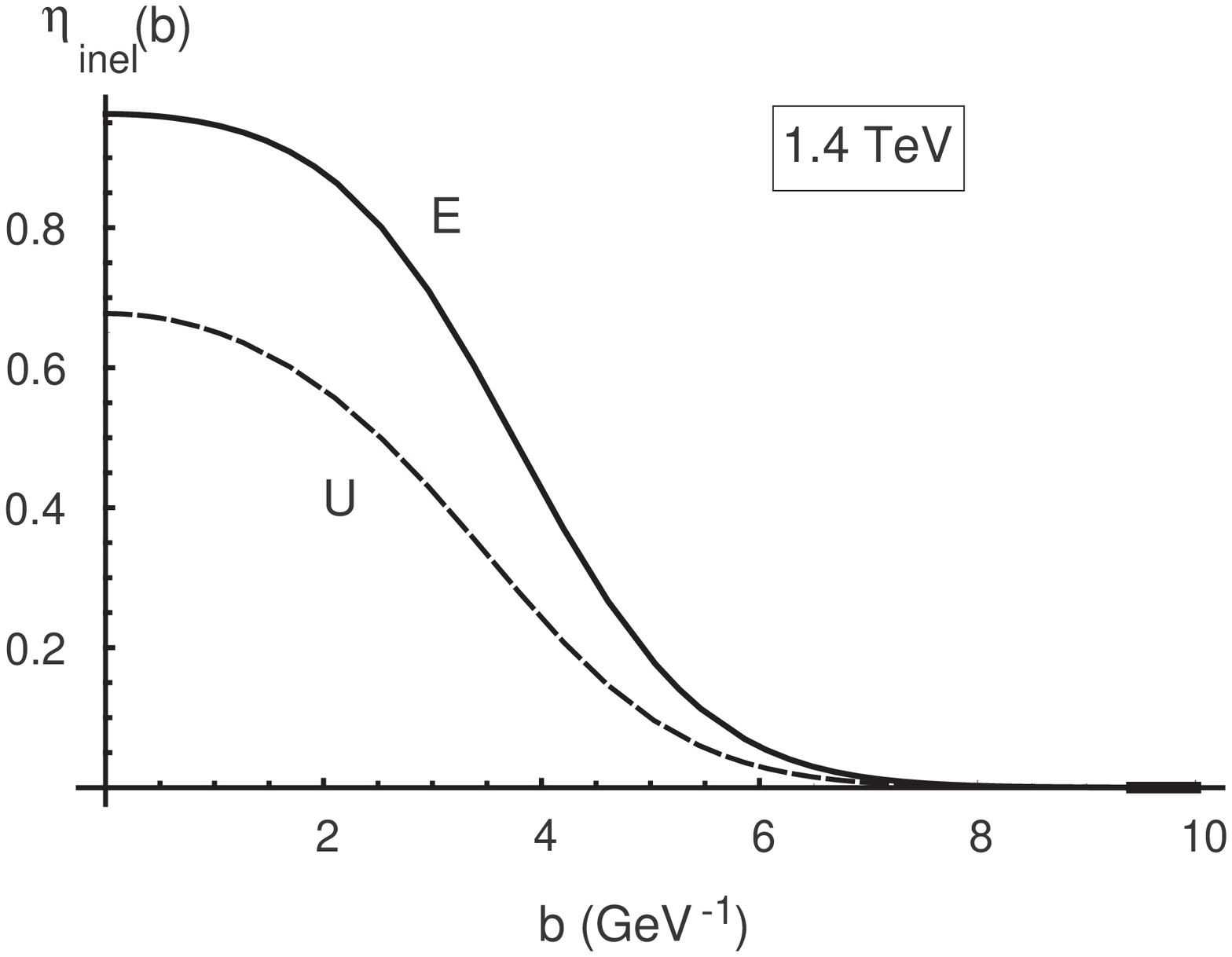}
\epsfysize=50mm\epsfbox{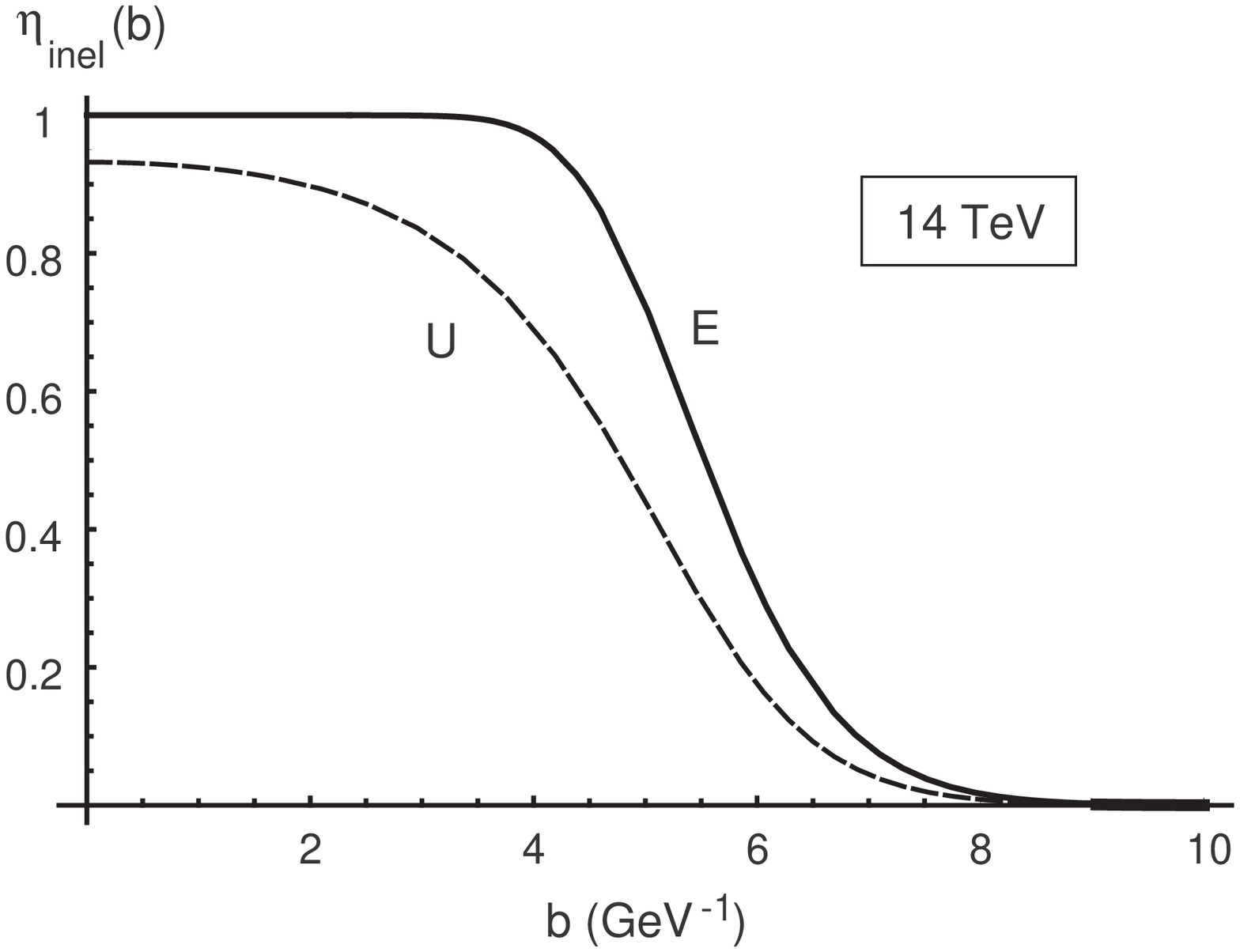}
\end{center}
\caption{Inelastic overlap function with eikonal (plane line) and
 $U$ matrix (dashed line) unitarization, at $\sqrt{s}=1.4 \ $TeV 
(left) and $\sqrt{s}=14 \ $TeV  (right), for a Born amplitude
$\chi(s,b) = 0.005 \ s^{0.4} \  exp(-b^2/3^2) $.  }
\end{figure}

Now let us compare directly the different forms
of  unitarization schemes.  In Ref. \cite{bg},
the eikonal and $U$-matrix phases are compared from
Eqs. (\ref{eik1}) and (\ref{U-mt}). Assuming that the two unitarised amplitudes are equal, one obtains
\begin{eqnarray}
  \chi_u(s,b) =2 \ \tanh\left[\frac{1}{2} \chi_e(s,b)\right]. \label{tanh}
\end{eqnarray}
  It is more convenient to analyze the inverse relation
\begin{eqnarray}
  \chi_e(s,b) = \log\left[\frac{1+ \chi_u(s,b)/2}{1- \chi_u(s,b)/2}\right].
  \label{chieu}
\end{eqnarray}
  Both dependencies are shown in  Fig. 4.
 At low energy, we have approximately the same size for the two phases.
  But at high energy, when  $\chi_e (s,b)$ goes to  infinity,
  $\chi_u (s,b) \rightarrow 2$. Hence the energy dependences of these 
phases are very different.

At asymptotic energy, if $\chi_u(s,b)$ grows
like the standard Born amplitude, $\chi_e(s,b)$ will tend to 
$i \pi $, as can be seen from Eq. (\ref{tanh}).
So the high-energy limit of $U$-matrix unitarisation corresponds 
to  the low-energy of the eikonal amplitude.
Indeed, for the standard eikonal at low energy we obtain
the saturation of the maximum unitarity bound  $=2$ in our
normalization (see Fig. 5a) in the case of a purely real
scattering amplitude. Larger values of the real
part of the scattering amplitude lead to a faster decrease of
this bound with some oscillations (see Fig. 5b).

If we compare the phases of the extended and standard eikonal 
unitarization schemes, we obtain a similar result:
the phase of the extended eikonal will also be bounded
   \begin{eqnarray}
  \chi_{re}(s,b) = -\log\left[\frac{1}{2}(1+e^{-\chi_e(s,b)})\right].
  \label{chieur}
\end{eqnarray}
Hence at high energies , when $\chi(s,b) \rightarrow \infty$,
$\chi_{e}(s,b)$ will be  bounded by  $0.693$.

\begin{figure}[!ht]

\centerline{
\epsfysize=35mm\epsfbox{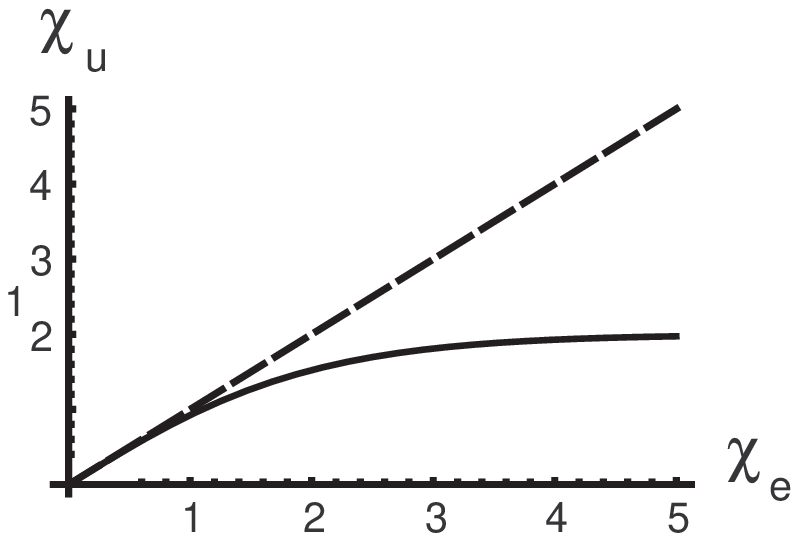}
\epsfysize=35mm\epsfbox{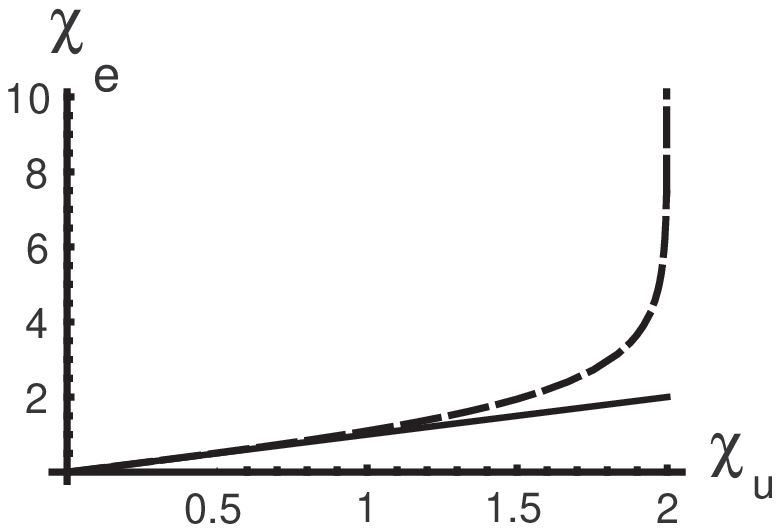}
}

\caption{The correlation of $\chi_{e}(s,b)$ and $\chi_{u}(s,b)$
  obtained from the eqs. (\ref{eik1}) and (\ref{U-mt})
 (dashed line is $\chi_{u}(s,b)$, plane line is $\chi_{e}(s,b)$
 }
\end{figure}

\begin{figure}[!ht]
\begin{flushleft}
\epsfysize=50mm
\epsfbox{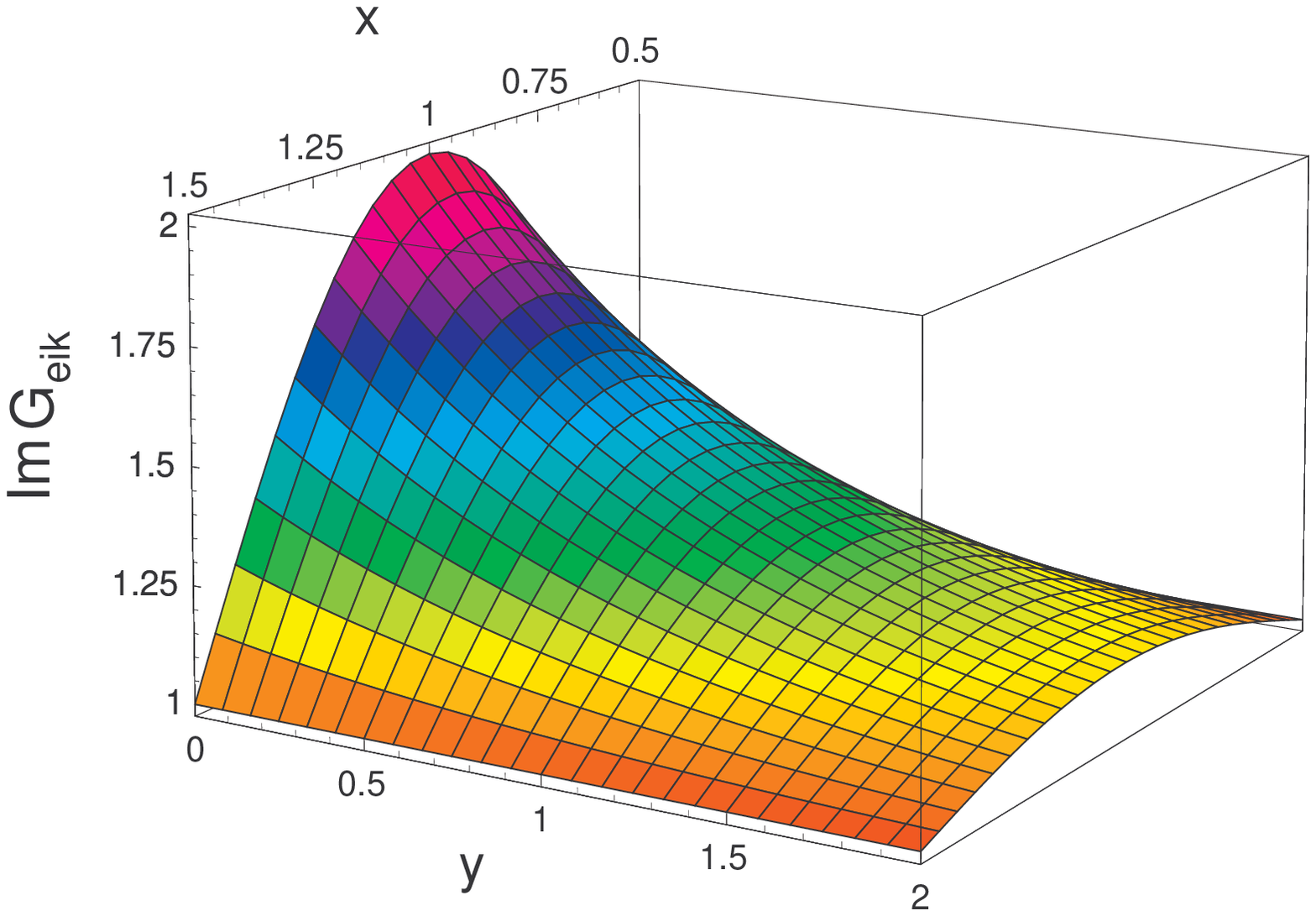}
\end{flushleft}
\vspace{-5.5cm}
\begin{flushright}
\epsfysize=50mm
\epsfbox{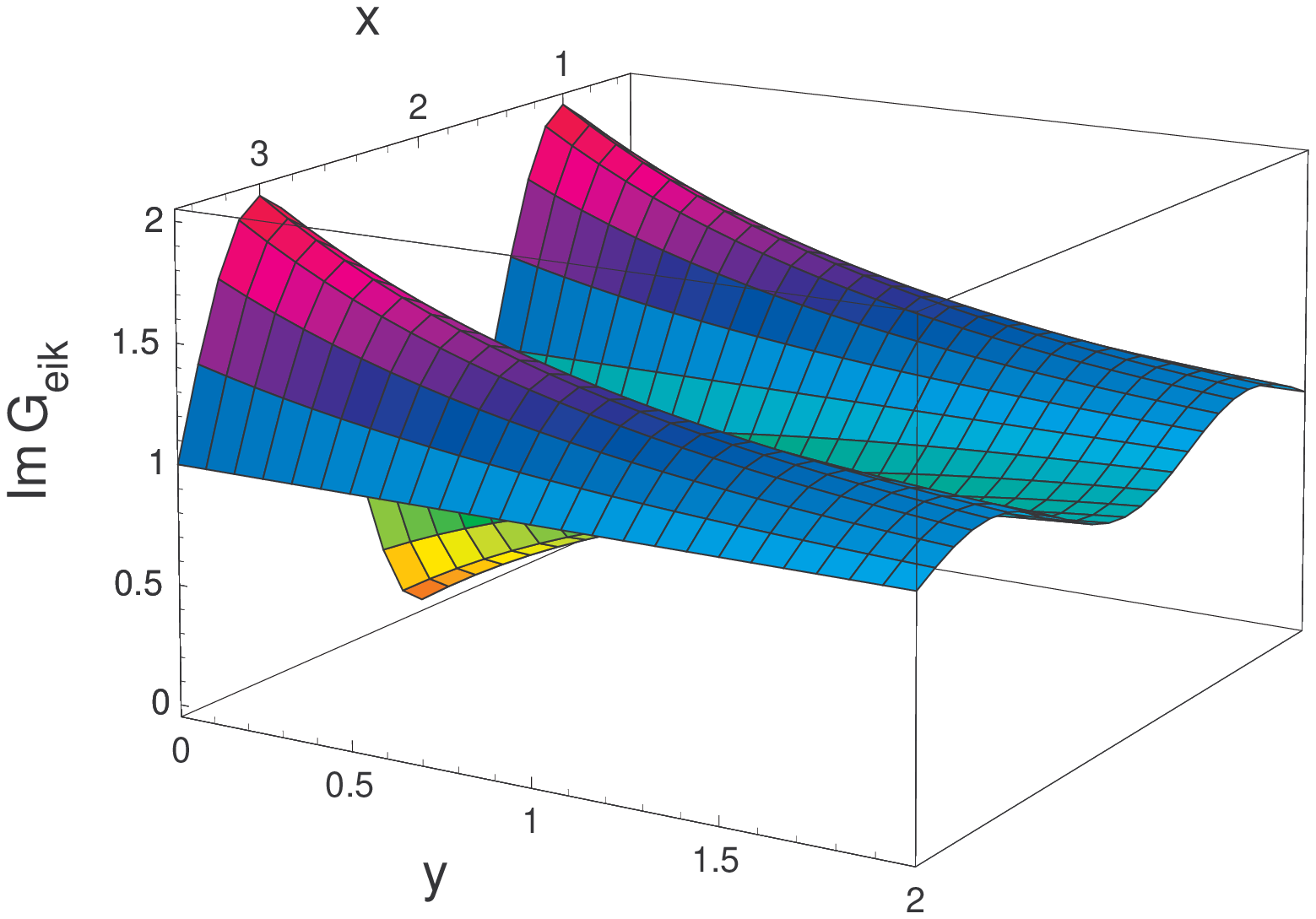}
\end{flushright}
\caption{The imaginary part of the unitarised eikonal in the region of 
small (a: left panel) and large (b: right panel) real part ($x$ axis) of 
the scattering amplitude.
 }
\end{figure}

     However, if we take the $U$ matrix in the form (\ref{Ue}),
  the relation between phases will be
\begin{eqnarray}
  \chi_e(s,b) = \log\left[1+ \chi_u(s,b)\right]  \label{eikm1}
\end{eqnarray}
whose inverse relation is
\begin{eqnarray}
  \chi_u(s,b) = e^{\chi_e(s,b)}\left[1 - e^{\chi_e(s,b)}\right]  
\label{umm1}
\end{eqnarray}
 In this case both phases  can vary from zero to infinity.

    \section{Conclusion}
From the above analysis we are led to conclude that the unusual properties
of the $U_T$-matrix unitarization are not connected with its specific form
as a ratio of polynomials. The exponential form can have the  same 
properties,  including the antishadowing regime and the unusual  asymptotic
ratio of   $\sigma_{el}$ to  $\sigma_{tot}$.
These unitarizations do not have the BDL at high energies,
and do not change significantly the  behavior of the scattering 
amplitude at LHC energies,
but they lead to a  faster growth of $\sigma_{tot}$.
Conversely, we have shown that an extended  $U$-matrix unitarization has 
the same properties as the standard eikonal unitarization.

The analysis of  the relation between the phases of the standard eikonal 
and of the $U_T$ matrix shows that if the phase of the $U_T$ matrix 
unitarization is bounded, both unitarization schemes will  have the same
properties. This would imply either that these unitarizations are valid 
only at low  energies, or that their Born term must profoundly change
at  high energies and be bounded.

Up to now, we have not found any decisive argument to forbid
unitarization approaches like the $U_T$ matrix. But the predictions 
for the total cross sections in these two approaches of the unitarization
have large differences (see, for example \cite{CS-BDL06,TTst}) for the LHC 
energy region, so that we hope that future experimental data will give us 
the true answer.
\section*{A note on the real part}
 
In a recent comment \cite{troshin}, Troshin has pointed out that the extended eikonal which we propose here unitarises only purely-imaginary amplitudes. We indeed considered
here the eikonalised amplitude
\begin{equation}
 f(s,b)=iG(s,b)  =  ik[1 \ - \ e^{i  (\phi(s,b)/k)}],
\label{eikext}
\end{equation}
with $k=2$, and $\Im m\phi=\chi$. As is well known \cite{Martynov1}, this maps the 
complex half plane  $\Im m \phi\geq 0$
to twice the unitarity circle, and maps $\phi=i\infty$ to 2$i$. So it cannot be considered a generic unitarisation scheme. 

However, we are not interested
here to map the whole half-plane to the circle: indeed, we only want a scheme that 
unitarises at high energy. We see no reason why an arbitrary amplitude should always be
unitarised: surely, the multiple exchanges depend on the underlying theory, and a scheme
that would work in one theory has no reason to work in another. Hence, we are interested
in unitarising high-energy hadronic amplitudes. These are dominated by their imaginary part at
high energy, so that the scheme we propose here produces very small deviations from unitarity at high energy for amplitudes which fit the lower-energy data. 
The unitarity condition $|f-i|^2\leq 1$ can indeed be rewritten
\begin{equation}
(k-2)+k\rho^{2}\leq 2(k-1)\rho \cos(\Re e\phi/k)
\end{equation}
with $\rho=e^{-\Im m \phi/k}$.
When one includes a real part in the Born term, we see that eventually $k=2$ will
lead to a contradiction as the cosine term goes to zero. However, because the latter
is suppressed by $\rho$, we also see that values of $k$ very close to 2 will 
lead to a proper solution. In the case of the 2-pomeron model of \cite{CS-BDL06},
we find that values $k=1.95$ do not lead to a violation of unitarity at high energy,
and keep the properties of antishadowing described in this paper.
\vspace{0.5cm}

\noindent{\small The authors would like to thank  for helpful discussions
  J. Fischer and E. Martynov, and acknowledge extended discussions with S.M. Troshin.
 O.S. gratefully acknowledges financial support
  from FNRS and would like to thank the  University of Li\`{e}ge
  where part of this work was done.
    }

\end{document}